\documentclass[12pt]{article}

\usepackage{amsmath, amssymb, slashed, epsf, color, graphicx, subfigure}

\begin{document}

\begin{titlepage}
\begin{center}

\hfill IPMU-14-0304 \\
\hfill DESY  14-168 \\
\hfill \today

\vspace{1.5cm}
{\large\bf Mass of Decaying Wino from AMS-02 2014}

\vspace{2.0cm}
{\bf Masahiro Ibe}$^{(a, b)}$,
{\bf Shigeki Matsumoto}$^{(b)}$, \\
{\bf Satoshi Shirai}$^{(c)}$
and
{\bf Tsutomu T. Yanagida}$^{(b)}$ \\

\vspace{2.0cm}
{\it
$^{(a)}${Institute for Cosmic Ray Research (ICRR), Theory Group, \\
University of Tokyo, Kashiwa, Chiba 277-8568, Japan} \\
$^{(b)}${Kavli Institute for the Physics and Mathematics of the Universe (IPMU),\\
University of Tokyo, Kashiwa, Chiba 277-8568, Japan} \\
$^{(c)}${Deutsches Elektronen-Synchrotron (DESY), 22607 Hamburg, Germany} \\
}

\vspace{2.5cm}
\abstract{
We revisit the decaying wino dark matter scenario in the light of the updated positron fraction, electron and positron fluxes in cosmic ray recently reported by the AMS-02 collaboration. We show the AMS-02 results favor the mass of the wino dark matter at around a few TeV, which is consistent with the prediction on the wino mass in the pure gravity mediation model.}

\end{center}
\end{titlepage}
\setcounter{footnote}{0}

\section{Introduction}

The AMS-02 collaboration has recently updated the positron fraction, the electron flux, and the positron flux in cosmic ray\,\cite{Accardo:2014lma, Aguilar:2014mma}, which consistently show anomalous excesses over the expectation based on the conventional comic-ray propagation model. Of particular interest of these new results is that the positron fraction is no longer increasing with energy above around 200\,GeV. Furthermore, the positron fraction and flux look to peak at around 300\,GeV. If the anomalous excesses come from the decaying dark matter, such ``peak" structures give constraints on the dark matter mass. In this letter, we revisit the decaying wino dark matter scenario\,\cite{Shirai:2009fq} to account for the anomalous excesses, and derive the constraint on the decaying wino mass along the line of the analysis of our previous paper\,\cite{Ibe:2013jya}.

From phenomenological viewpoint, the wino-like dark matter is a good candidate for weakly interacting massive particle (WIMP) that evades sever limits from direct detection searches\,\cite{Hisano:2012wm}. On the other hand, the dark matter predicts strong signals in indirect detection searches utilizing e.g. gamma-ray observations, for its annihilation cross section is boosted by the Sommerfeld enhancement\,\cite{Hisano:2003ec, Hisano:2004ds}. Though the strength of the signal is still below current experimental limits due to large astrophysical ambiguities, they are expected to be detected in near future\,\cite{Bhattacherjee:2014dya}.

From theoretical viewpoint, the wino-like dark matter is realized in a wide class of supersymmetric standard models when gaugino masses are dominated by the anomaly mediated supersymmetry breaking contributions\,\cite{AMSB}. Models with anomaly mediated gaugino mass are now highly motivated since they provide a good dark matter candidate (i.e. the wino) while explaining the observed Higgs boson mass about 126\,GeV\,\cite{OYY} in conjunction with the high scale supersymmetry breaking where the gravitino and the sfermion masses are 
in ${\cal O}(100-1000)$\,TeV range\,\cite{Wells:2004di}. Such models are, for example, realized as the models of pure gravity mediation (PGM)\,\cite{PGMs, PGM2, Evans:2013lpa}, the models with strong moduli stabilization\,\cite{Strong Moduli}, the spread supersymmetry\,\cite{spread}, and the minimal split supersymmetry\,\cite{Split SUSY}. As we will show, the recent observations of AMS-02 suggest the decaying wino mass is at around a few TeV, which is consistent with the prediction on the wino mass in this class of models.

\section{Decaying wino in the PGM model}

Let us briefly summarize the decaying wino dark matter scenario in the pure gravity mediation model. In the model, gaugino masses are dominated by the one-loop anomaly mediated contributions\,\cite{AMSB}, and the neutral wino becomes the lightest supersymmetric (LSP). For derivation of the anomaly mediated gaugino masses in superspace formalism of supergravity, refer the papers\,\cite{Bagger:1999rd, D'Eramo:2013mya, Harigaya:2014sfa}. The Higgsino mass term is, on the other hand, generated through tree-level interactions to the $R$-symmetry breaking sector\,\cite{Casas:1992mk} (the generalized Giudice-Masiero mechanism\,\cite{Giudice:1988yz}), which leads to the Higgsinos mass much larger than the gaugino masses. With such a large Higgsino mass term, the mixing between the wino and the bino is highly suppressed. The PGM model therefore predicts the almost pure neutral wino as the LSP which is a good candidate for WIMP dark matter.

The wino dark matter is produced thermally in the early universe and non-thermally by the decay of the gravitino in the late universe before the big bang nucleosynthesis (BBN) starts. Putting these contributions together, the wino mass turns out to be lighter than about 3\,TeV in order to be consistent with the observed dark matter density\,\cite{Ade:2013zuv}. The wino mass of around 3\,TeV is particularly interesting, because the dark matter density is explained solely by its thermal relic density\,\cite{Hisano:2006nn}. The wino dark matter lighter than 3\,TeV can also provide the correct relic density when the non-thermal production dominates,\cite{ Gherghetta:1999sw, Moroi:1999zb, PGMs}. In particular, the wino mass below about $1$--$1.5$\,TeV is interesting, because such a lighter wino dark matter is easily consistent with the traditional thermal leptogenesis scenario\,\cite{leptogenesis}.

The wino dark matter is not necessarily to be absolutely stable and may decay into standard model particles when the $R$-parity is slightly violated. In this letter, we consider the decay caused by $R$-parity violating interactions,
\begin{eqnarray}
{\cal W}_{\slashed{R}} = \lambda_{ijk} \, L_iL_jE^c_k\ ,
\label{eq: LLE}
\end{eqnarray}
among various possibilities to violate the $R$-parity. Here, the indices denote the generation of leptons, and $\lambda$'s are tiny coupling constants. The decaying wino dark matter via the $LLE^c$ interactions is free from the constraint from cosmic-ray anti-proton observations.\footnote{We presume the absence of other $R$-parity violating operators. As discusses in reference\,\cite{Ibe:2013jya}, it is possible to generate only $LLE^c$ operators in a grand unified theory consistent way.} Constraints from gamma-ray observations are also much milder than the case of models with $R$-parity violation by $LH_u$\,\cite{Shirai:2009fq}. Through the $R$-parity violating interactions in equation\,(\ref{eq: LLE}), the wino dark matter decays into three-body final states that are composed only of leptons (a pair of charged leptons and a neutrino). Its lifetime is estimated to be as follows\,\cite{Barbier:2004ez}:
\begin{equation}
\tau_{\rm wino} \sim
10^{27}\,(\lambda/10^{-19})^{-2}\,(m_{\rm wino}/1~{\rm TeV})^{-5}\,
(m_{\tilde L}/10^3~{\rm TeV})^4\,{\rm sec.}\ .
\end{equation}
Electron and positron cosmic rays from the decay reproduce the anomalous excesses of AMS-02 for $\tau_{\rm wino} = O(10^{26-27})$\,sec, as will be seen in the next subsection.

\section{Wino mass from AMS-02 2014}
\label{sec: results}

The procedure to calculate the electron and positron fluxes for signal and background is essentially the same as the one adopted in our previous paper\,\cite{Ibe:2013jya}. We made several assumptions in the procedure, and those are listed below in order.

\begin{itemize}

\item
The decay of the wino dark matter is described by the interaction $L_i L_j E^c_k$. Primary $e^+$ and $e^-$ spectra from the decay is obtained assuming the left-handed slepton $\tilde{L}_i$ is enough lighter than others and no flavor violation exists on couplings between wino and (s)leptons. We have used Pythia\,8 \,\cite{Sjostrand:2007gs} for the spectra with a slight modification for a polarized lepton decay. The dark matter mass density of our galaxy is assumed to follow the NFW profile\,\cite{Navarro:1996gj} with profile parameters $\rho_\odot = 0.4~{\rm GeV/cm^3}$ (the local halo density), $r_c = 20~{\rm kpc}$ (the core radius), and $r_\odot = 8.5~{\rm kpc}$ (the distance between our solar system and the galactic center). Propagations of the electrons and positrons in our galaxy are considered using the diffusion equation of the so-called MED model\,\cite{Delahaye:2007fr}.

\item
For astrophysical backgrounds against the signals, we have adopted a similar method developed in reference\,\cite{Cirelli:2008pk}. Using parameters $A^\pm$ and $p^\pm$, background fluxes are parameterized as $\Phi^{e^\pm}_{\rm BG}(E) = A^\pm \, E^{p^\pm} \, \Phi^{e^\pm}_{\rm ref}(E)$. Here, $\Phi^{e^\pm}_{\rm ref}(E)$ are reference background fluxes obtained by GALPROP\,\cite{Strong:1998pw} with the electron injection index being $-2.66$. Effect of the solar modulation is also considered by the force-field method\,\cite{Gleeson:1968zza} in both signal and background calculations.

\item
Above electron and positron fluxes are fitted to the latest AMS-02 data of the positron fraction\,\cite{Accardo:2014lma} and the electron flux\,\cite{Aguilar:2014mma}. Following six parameters, the background parameters ($A^\pm$, $p^\pm$) and the force-field potentials for electrons and positrons ($\phi^\pm$), are varied to maximize the likelihood function of the fitting for each wino mass and lifetime ($m_{\rm wino}$ and $\tau_{\rm wino}$) in the ranges of $A^\pm \in [0, \infty]$, $p^\pm \in [-0.5, 0.5]$, and $\phi^\pm \in [0, 1]$ GV, respectively. The fitting has been performed in the energy range of $E > 5$~GeV for the positron fraction and $>10$ GeV for the electron flux to suppress the effect of the solar modulation.

\end{itemize}

\begin{figure}[h!]
\begin{center}
\subfigure[$L_1 L_2 E^c_k$]{
\includegraphics[clip, width = 0.45 \textwidth]{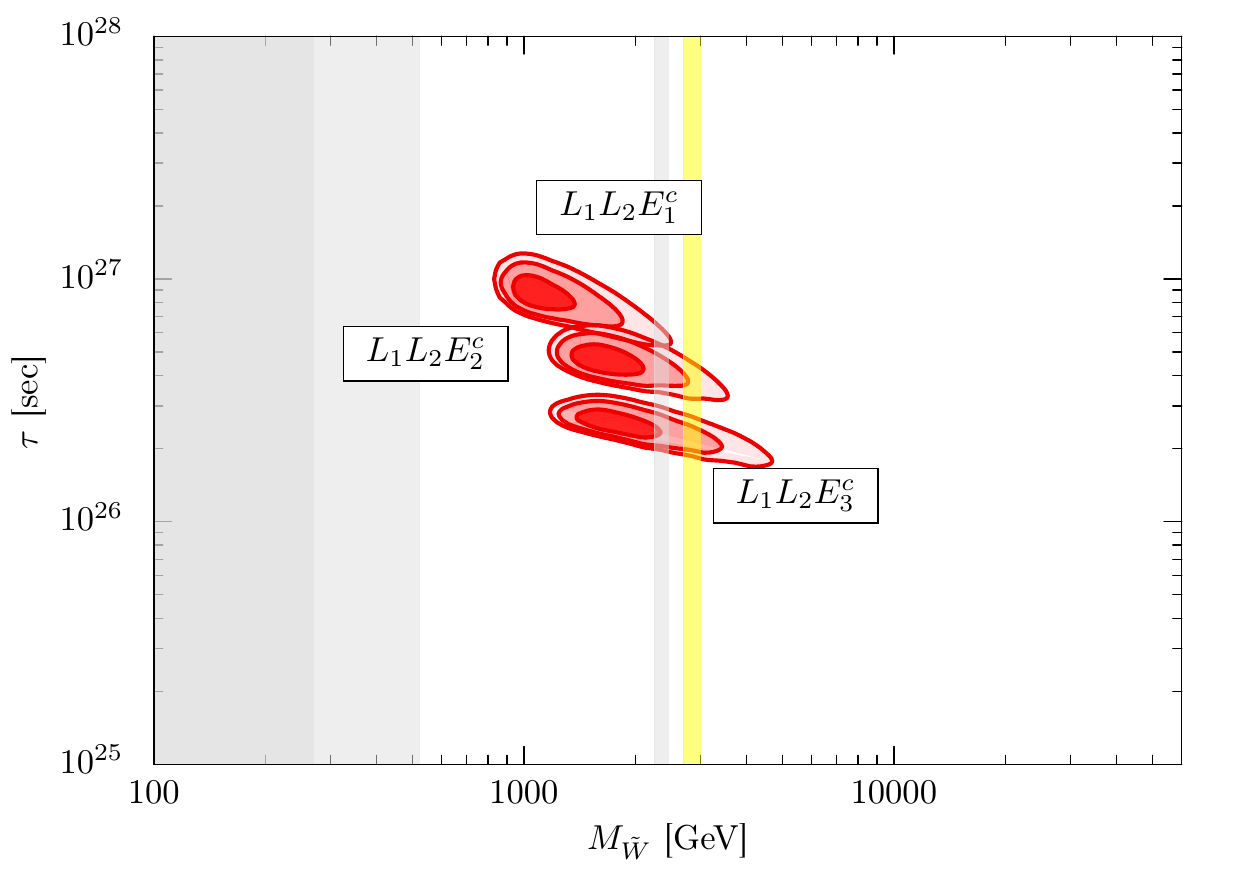}
}
\subfigure[$L_3 L_1 E^c_k$]{
\includegraphics[clip, width = 0.45 \textwidth]{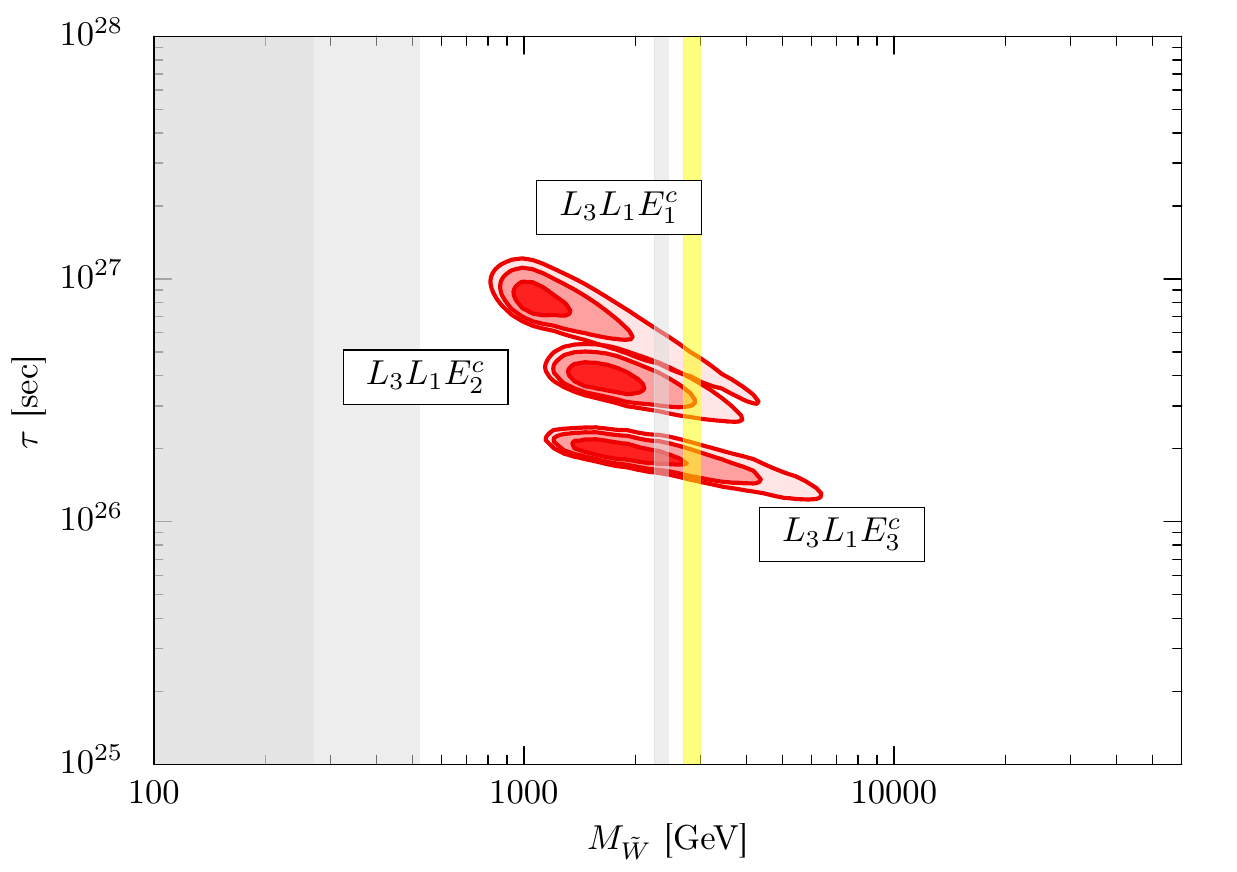}
}
\subfigure[$L_3 L_2 E^c_k$]{
\includegraphics[clip, width = 0.45 \textwidth]{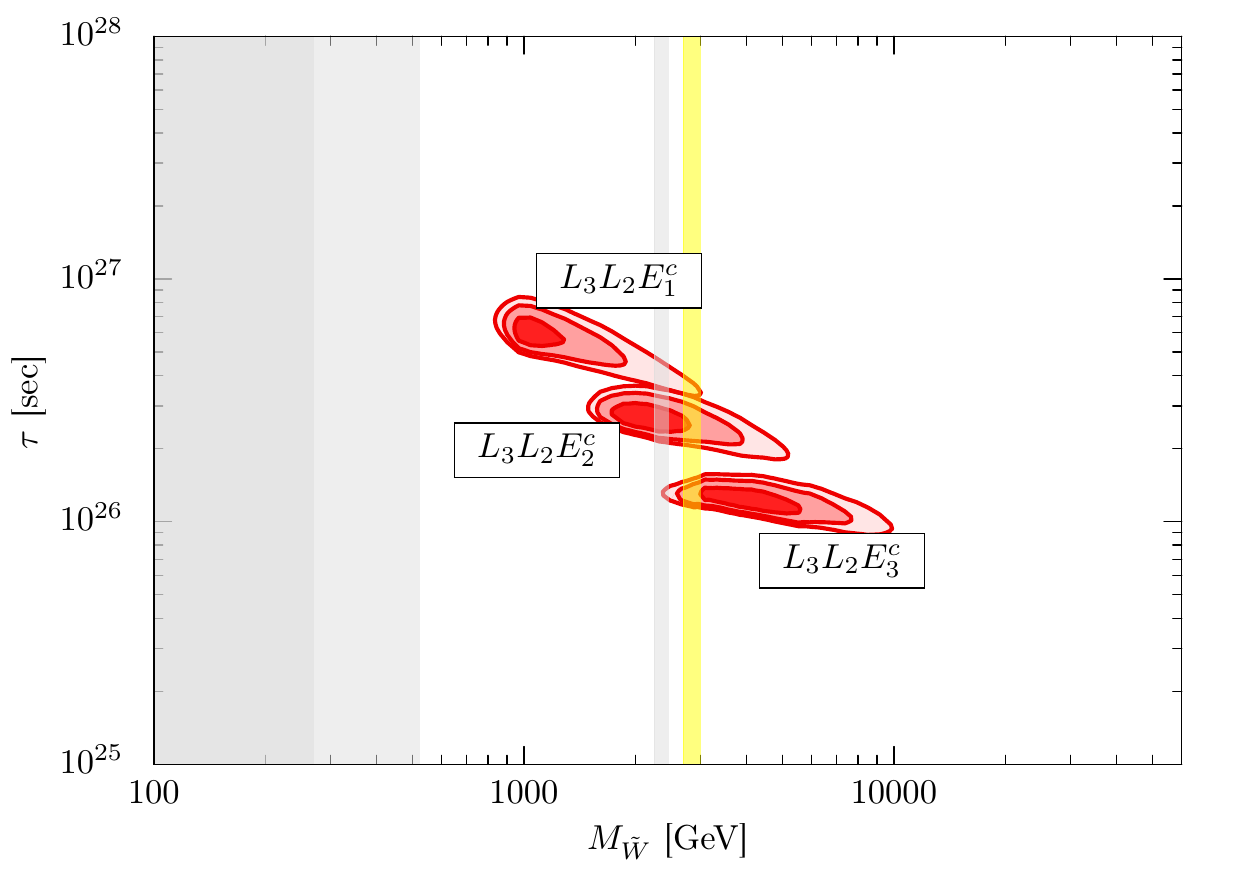}
}
\subfigure[$L_3 L_1 E^c_2$ positron fraction]{
\includegraphics[clip, width = 0.45 \textwidth]{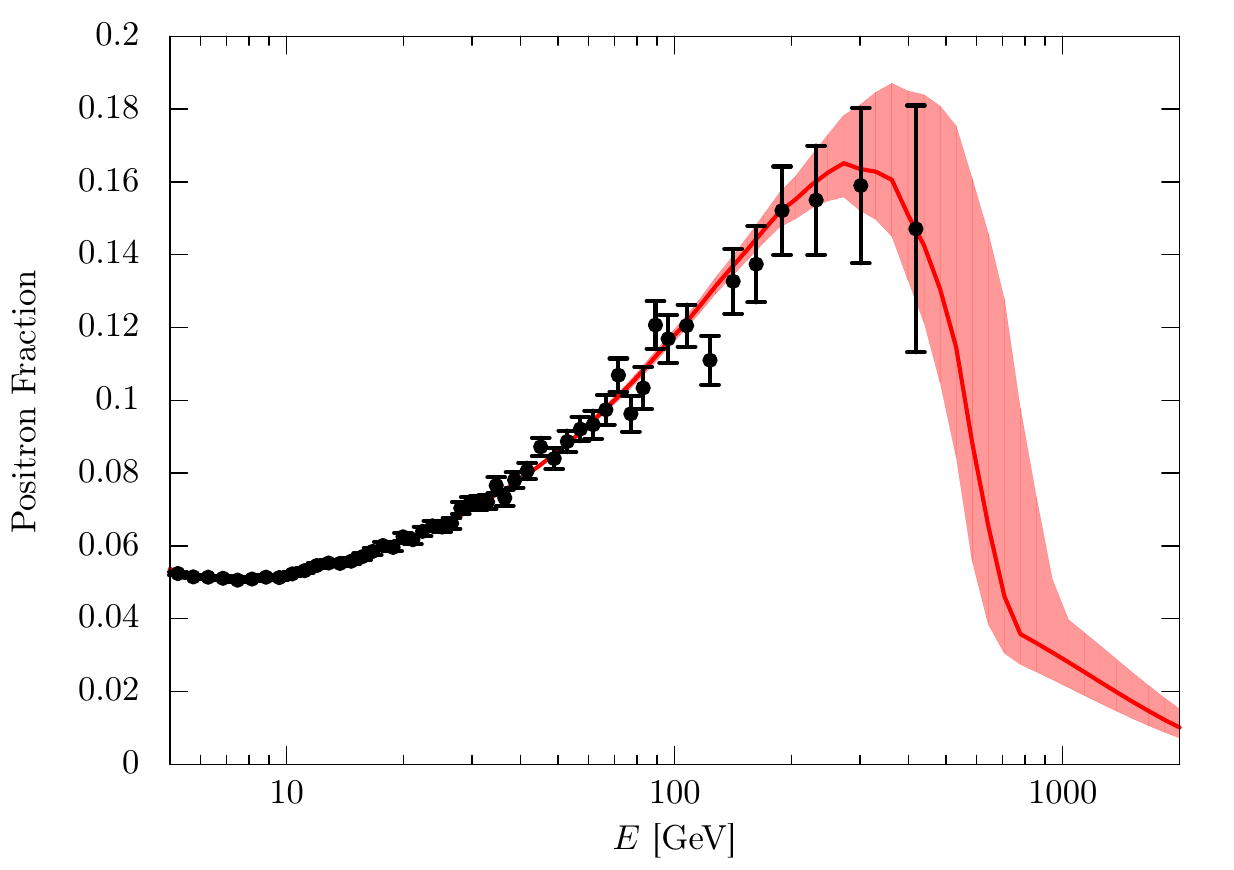}
}
\subfigure[$L_3 L_1 E^c_2$ electron flux]{
\includegraphics[clip, width = 0.45 \textwidth]{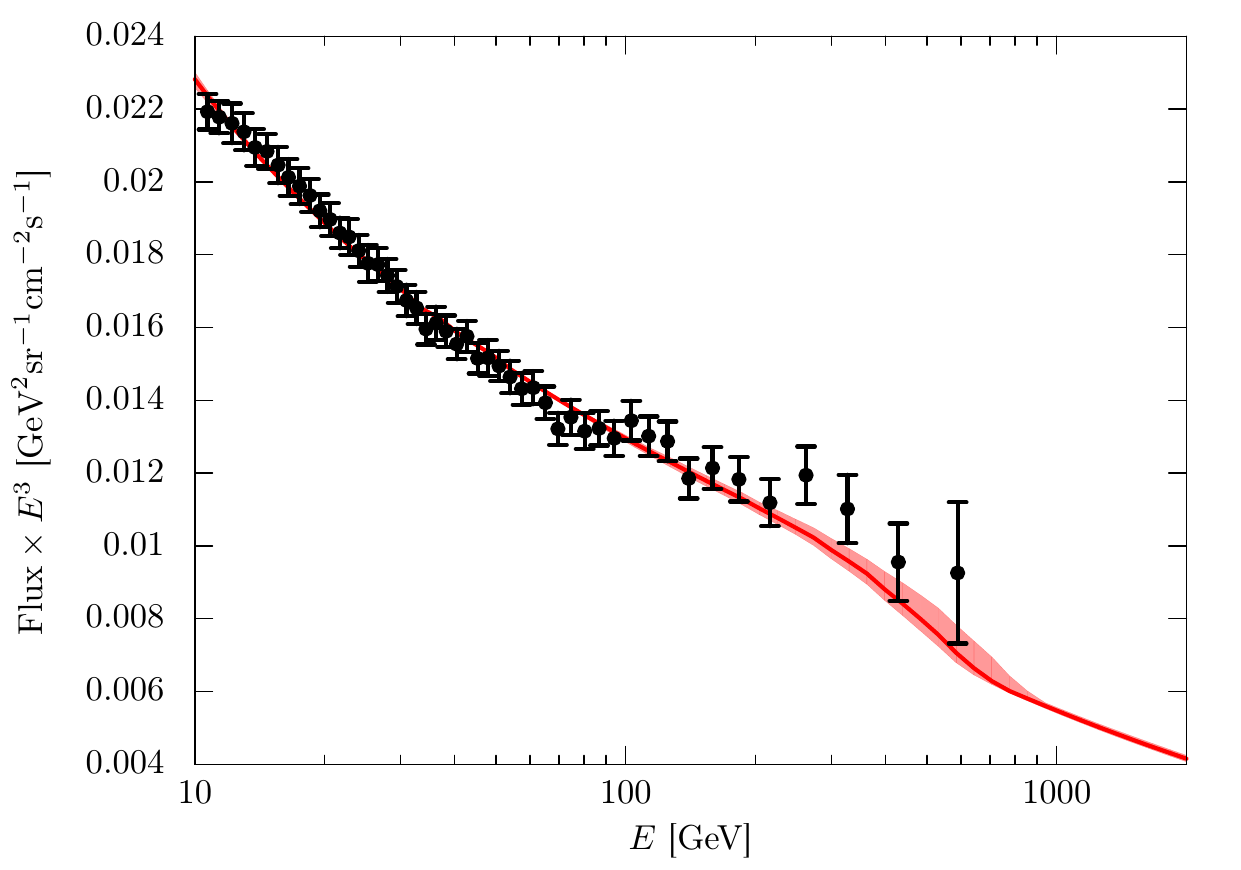}
}
\subfigure[$L_3 L_1 E^c_2$ positron flux]{
\includegraphics[clip, width = 0.45 \textwidth]{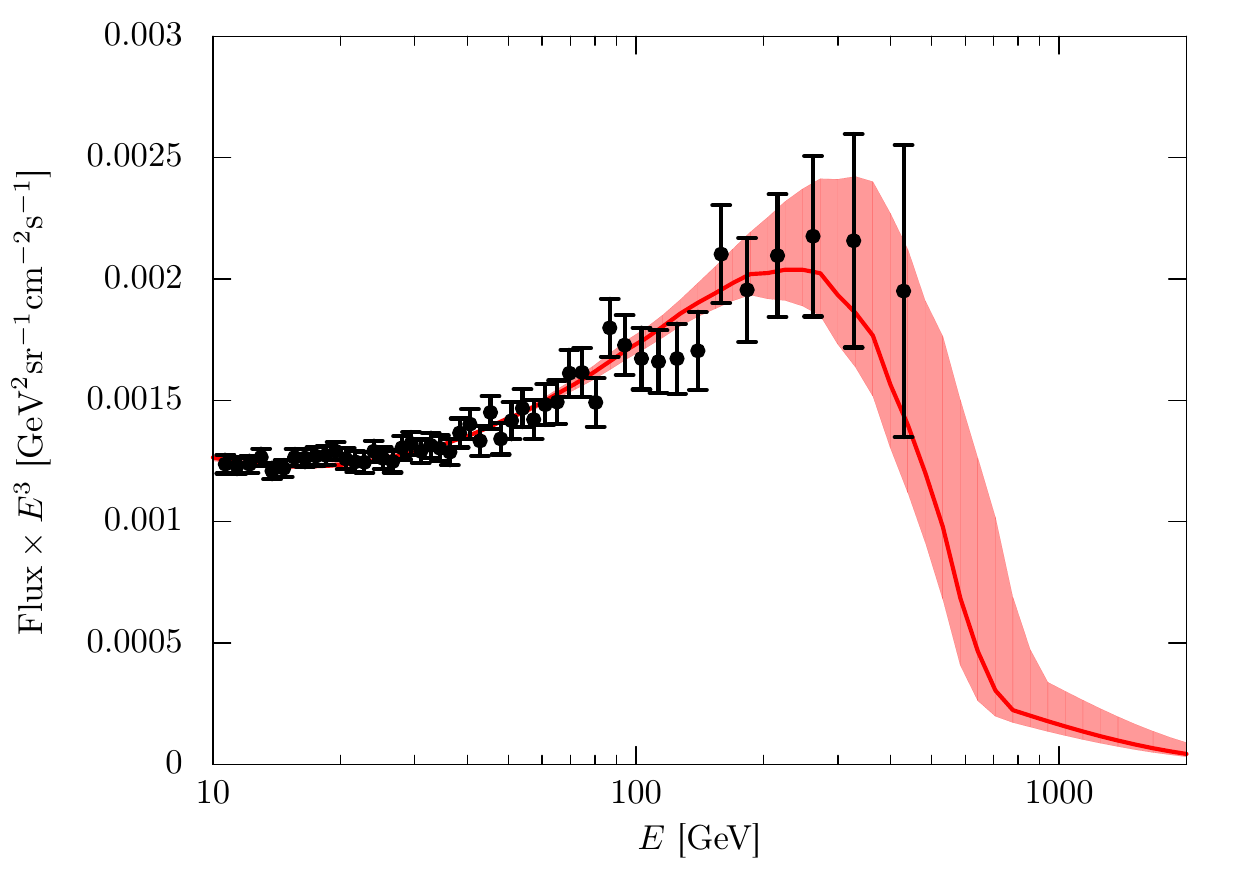}
}
\caption{\small {\bf Upper three panels}: Contour lines of 68th, 95th, and 99th percentile of the likelihood function (the chi-squared distribution) for the wino decays through the interactions $L_1 L_2 E^c_i$ (top-left panel), $L_3 L_1 E^c_i$ (top-right panel), and $L_3 L_2 E^c_i$ (middle left panel), where $i = 1, 2,\,{\rm and}\,3$. See text for gray and yellow shaded regions. {\bf Lower three panels}: The positron fraction (middle-right panel), the electron flux (bottom-left panel), and the positron flux (bottom-right panel) with the latest AMS-02 data for the decay through the interaction $L_3 L_1 E^c_2$. See text for red solid lines and red shaded regions.}
\label{fig: results}
\end{center}
\end{figure}

Fitting results are depicted in upper three panels of figure~\ref{fig: results} as contour lines of 68th, 95th, and 99th percentile of the chi-squared distribution for two degrees of freedom. The results for the wino decays caused by the interactions $L_1 L_2 E^c_i$, $L_3 L_1 E^c_i$, and $L_3 L_2 E^c_i$ ($i = 1, 2,\,{\rm and}\,3$) are shown in top-left, top-right, and middle-left panels, respectively. As a reference, the wino mass favored by the thermal WIMP scenario is shown as a light yellow bar. In lower three panels of the figure, as an example, the positron fraction (middle-right panel), the electron flux (bottom-left panel), and the positron flux (bottom-right panel) are shown with the latest AMS-02 data for the decay caused by the interaction $L_3 L_2 E^c_1$. The red solid lines in these plots are from the best-fit parameters of $m_{\rm wino}$ and $\tau_{\rm wino}$, while red shaded regions are obtained by the parameters within 68th percentile of the chi-squared distribution.

Constraints on the wino mass from other experiments are also shown in the upper three panels as regions shaded by grays: The dark gray region in each panel is from the disappearing charged track search at the Large Hadron Collider experiment~\cite{ Aad:2013yna, Ibe:2012sx}. The light gray region is from the Fermi-LAT experiment~\cite{Ackermann:2013yva}, which is obtained by observing gamma-rays from the wino dark matter annihilation at classical dwarf spheroidal galaxies. This observation is known to give the most robust limit on the wino mass among various indirect detection searches of dark matter~\cite{ Bhattacherjee:2014dya}. In addition, there is another constraint on $m_{\rm wino}$ and $\tau_{\rm wino}$, which is again from the Fermi-LAT experiment observing diffuse gamma-rays from the wino decay. The observation gives the constraint as $\tau_{\rm wino} \gtrsim 10^{26}$\,sec in the region of $m_{\rm wino} \sim 1$\,TeV when we use officially published data of the Fermi-LAT experiment~\cite{Abdo:2010nz, Ackermann:2012qk}. We did not explicitly show the constraint on the panels to avoid making the figure busy.%
\footnote{
For more details of the constraint, see our previous paper~\cite{Ibe:2013jya}.
}

As can be seen from the figure, the decaying wino dark matter with $LLE^c$ interactions is indeed very consistent with the latest AMS-02 data. In particular, the wino mass favored by the data is always within a few TeV region irrespective to the lepton flavor structure of the interaction $L_i L_j E^c_k$, which is nothing but the region predicted by the pure gravity mediation model. The limit from the diffuse gamma-ray observation seems to start excluding the favored parameter region when the wino decays mainly into tau leptons, such as the decays caused by the $L_3 L_1 E^c_3$ and $L_3 L_3 E^c_3$ interactions. On the other hand, the decays mainly into first and second generation leptons are still away from the limit. For the sake of convenience, we have also estimated the uncertainty associated with electron and positron propagations in our galaxy using the diffusion equations of the so-called M1 and M2 models. The uncertainty turns out not to change the result drastically.

\section{Summary and discussions}

We have revisited the decaying wino dark matter scenario in the light of the updated positron fraction, electron flux, and positron flux in cosmic ray reported by the AMS-02 collaboration. The AMS-02 data can be well explained by the almost pure wino dark matter of its mass around a few TeV and its decay described by the $R$-parity violating $LLE^c$ interactions. Such a dark matter, in particular a few TeV range of the wino mass, is consistent with the pure gravity mediation model very well.

The origin of the anomalous excess reported by the AMS-02 collaboration is still unknown: Both dark matter interpretation and astrophysical interpretation such as pulsar activities nearby us may be still possible (for recent discussions 
see \cite{Linden:2013mqa,Yin:2013vaa,Lin:2014vja} 
and references therein). Future observations of extragalactic diffuse gamma-rays caused by the wino decay and those of gamma-rays caused by the wino dark matter annihilation at dwarf spheroidal galaxies will be important to convince us that the wino dark matter is really the origin.

If the $R$-parity breaking operators such as $U^c U^c D^c$ or $Q D^c L$ are not completely suppressed, the wino decay may have the hadronic modes.
If it is the case, we may have antiproton excess in the cosmic ray.

\vspace{1.0cm}

\noindent
{\bf Acknowledgments}
\vspace{0.1cm}\\
\noindent
This work is supported by the Grant-in-Aid for Scientific research from the Ministry of Education, Science, Sports, and Culture (MEXT), Japan [No. 26104009/26287039 (M.I., S.M., T.T.Y.) and No. 24740151/25105011 (M. I.)], from the Japan Society for the Promotion of Science (JSPS) [No. 26287039 (M.I.)], as well as by the World Premier International Research Center Initiative (WPI), MEXT, Japan.


\end{document}